# Single NanoParticle Photothermal Tracking (SNaPT)

# of 5 nm gold beads in live cells

David Lasne*, Gerhard A. Blab*, Stéphane Berciaud, Martin Heine[#], Laurent Groc[#], Daniel Choquet[#], Laurent Cognet, and Brahim Lounis

*Centre de Physique Moléculaire Optique et Hertzienne, CNRS (UMR 5798) and Université Bordeaux 1,*
*351 Cours de la Libération, 33405 Talence Cedex, France*
*[#] Physiologie Cellulaire de la Synapse, CNRS (UMR 5091) and Université Bordeaux 2*
*Institut François Magendie, 33077 Bordeaux Cedex, France*

\* These authors contributed equally to this work

Classification: Biological Science (Biophysics)

**Corresponding author:**

Prof. B. Lounis
Centre de Physique Moléculaire Optique et Hertzienne
CNRS UMR 5798 et Université Bordeaux 1
351 Cours de la Libération
33405 Talence, France
b.lounis@cpmoh.u-bordeaux1.fr

**Manuscript information :**

14 text pages (including title page, abstract, text, acknowledgements, references, figure legends),
5 pages of figures,
Supporting information
1 page
1 movie file.

**Word and character counts :** 122 words in the abstract, 28871 characters in paper

**Abstract**

*Tracking individual nano-objets in live cells during arbitrary long times is an ubiquitous need in modern biology. We present here a method for tracking individual 5 nm gold nanoparticles on live cells. It relies on the photothermal effect and the detection of the Laser Induced Scattering around a NanoAbsorber (LISNA). The key point for recording trajectories at video rate is the use of a triangulation procedure. The effectiveness of the method is tested against Single fluorescent Molecule Tracking in live COS7 cells on subsecond time scales. We further demonstrate recordings for several minutes of AMPA receptors trajectories on the plasma membrane of live neurons. SNaPT has the unique potential to record arbitrary long trajectory of membrane proteins using non-fluorescent nanometer sized labels.*

# Introduction

The movements of molecules in the plasma membrane of living cells are characterized by their diversities, both in the temporal and spatial domain. Membranes exhibit a constitutive complexity, consisting of several different lipids and a great variety of proteins with highly dynamic and compartmentalized spatial distributions. These molecules explore the plasma membrane in various lateral diffusion modes, and frequently interact with each other at specific locations in order to transmit information across the membrane, starting the cascades of specific signaling processes. Those molecular interactions are by nature heterogeneous making ensemble observations of these phenomena rather challenging. Single molecule detection has allowed the elimination of the implicit averaging of conventional optical observations, giving access to heterogeneity, dynamical fluctuations, lateral diffusion, reorientation, colocalization, and conformational changes at the molecular level. Until now, two main approaches have been used to track individual molecules in the plasma membrane of live cells, with distinct advantages and limitations. The first one, Single Particle Tracking (SPT), uses labels large enough to be detectable by conventional microscopes(1, 2) through Rayleigh intensity scattering (~ 40 nm gold particles or even larger latex beads). SPT permits to follow the movement of individual molecules for very long times and possibly at very fast imaging rates(3). SPT, for instance, revealed barriers set for diffusion by the cytoskeleton(4), and the diversity of lateral diffusion modes of receptor for neurotransmittors in live neurons(5, 6). However the main drawback is the size of the beads which might sterically hinder the interaction between the labeled molecules or alter their movements in confined environments such as synaptic clefts or endocytotic vesicles. Recently, a method which exploits the interference between a background reflection and the scattered field has been

developed(7). The detection of gold nanoparticles smaller than 20 nm has been achieved in vitro conditions(8, 9).

The second widely used technique, Single Molecule Tracking (SMT), uses fluorescent organic dyes(10, 11) or autofluorescent proteins(12). As these fluorophores are generally smaller than the target molecules it does not have the drawback of SPT mentioned previously. Applied to neurosciences, SMT has thus allowed to reveal the lateral diffusion of glutamate receptors (AMPA and NMDA) inside the synapses of live neurons(13, 14). The main limitation encountered in SMT studies is photobleaching which severely limits the observation times of a single fluorophore to typically less than one second in live cells.

An experimental technique combining the advantages of SPT and SMT, namely long observation times and small nanometer-sized labels, would thus have great potential. For biological questions, it would allow for recording the full history of proteins in cells including intermediate states even in highly confined regions (e.g. lipid rafts or membrane protein clusters, intracellular vesicles, synapses of neurons….).

For this purpose, semiconductor nanocrystals (such as CdSe/ZnS) have been used recently as fluorescent markers(15, 16). They are more photostable than organic dyes and autofluorescent proteins. Quantum dots have proven to be valuable tools for extended observation in living organisms(17, 18). However, biocompatible and functionalized nanocrystals are rather bulky labels of few tens of nanometers in diameter and they eventually bleach. In addition, their luminescence is subject to blinking(19) at all time scales(20) which renders observations of changes between different lateral diffusion modes of proteins difficult.

An alternative approach consists in developing sensitive optical systems for the detection of the absorption of nanoparticles (NPs). We have recently shown that the photothermal methods allow for the detection of individual non-luminescent nano-objects(21-23). Metal NPs are efficient light absorbers. The luminescence yield of these particles being extremely weak(24), almost all the absorbed energy is converted into heat. The increase of temperature induced by the absorption gives rise to a local variation of the refraction index. A Photothermal Interference Contrast technique was used to detect for the first time individual 5 nm gold NPs embedded in thin polymer films NPs(21) and to demonstrate the detection of single protein labeled with 10 nm gold NPs in fixed cells(22). Because of its limited sensitivity, this technique requires relatively high laser beam intensities restricting its use to fixed biological samples. Photothermal heterodyne imaging(23) is two orders of magnitude more sensitive than earlier methods. It allowed for the unprecedented detection of individual 1.4 nm gold NPs. It uses a combination of a time-modulated heating beam and a non-resonant probe beam. The heating induces a time-modulated variation of the refraction index around the absorbing NP. The interaction of the probe beam with this index profile produces scattered field with sidebands at the modulation frequency. The scattered field is then detected in the forward direction through its beatnote with the transmitted probe field which plays the role of a local oscillator akin a heterodyne technique(25). In the following, "Laser Induced Scattering around a NanoAbsorber" (LISNA) will be used to refer to this detection method. LISNA images are obtained by raster scanning of the samples by means of a piezoscanner stage (see fig1).

In this work, we used LISNA to perform Single nanoParticles Tracking (SNaPT) in live cells at video rate. For this purpose, we have developed a tracking strategy to record the 2D trajectories of individual moving 5 nm gold nanoparticules. The results obtained by SNaPT are compared to those obtained by SMT in a study of the lateral diffusion of individual metabotropic glutamate receptors on the plasma membrane of live COS7 cells. Furthermore, long trajectories (several minutes) of individual glutamate receptors (AMPA) recorded on the plasma membrane of live neurons are shown.

## Results and Discussion

While the sensitivity of LISNA is unparalleled, the imaging rate of this scanning method is a serious limitation. Recording the movement of membrane proteins on live cells requires fast acquisition rates (typically video rate or faster). The acquisition of several µm² images is however not necessary since the localisation of a single NP requires only a limited number of data points taken at well chosen position. For instance, tracking fluorescent objects with confocal microscopes has been proposed by using rotating illuminations(26) . It has been recently implemented in 2-photon microscopes and allowed to perform 3D tracks of 500 nm beads phagocytosed by live fibroblasts at video rate(27).

In order to study the lateral diffusion of membrane proteins in a plane, we present here an algorithm requiring 3 measurement points to localize a NP. The movements of the objects along the axial direction of the microscope are neglected since the cellular material will only consist on one layer of flat cultured cells. However, the algorithm can easily be extended to track 3D movements. We use 5nm gold particles as they can be imaged in live cells with high signal to noise ratios (SNR) at reasonable laser intensities (SNRs >30 for excitation intensities of ~400kW/cm², 5ms integration time per pixel).

*Description of the tracking method*

The spatial profile of the LISNA signal from an individual NP is given by the product of the intensity profiles of the heating and probe beams(25). It is well approximated by a Gaussian profile with a constant width, and can be defined by three parameters: the central position ($x_0$, $y_0$), and the peak signal $S_0$. By measuring the signal at three well defined positions around the tentative location of a NP, both its precise location in space ($x$, $y$) and its peak signal $S$ can be unequivocally retrieved.

In practice, after the acquisition of a white light and/or an epifluorescence image, a LISNA image is recorded in order to check gold-labeling density and specificity. Then a region of interest is chosen in which LISNA signals are taken at random positions. When a signal above a predefined threshold is obtained, three data point are taken around this position at the apices of an equilateral triangle. A first set of NP coordinates and peak signal is then calculated ($x_t$, $y_t$, $S_t$). If $S_t$ and the three measured signals are above a second predefined threshold, this procedure is repeated iteratively by recentering the equilateral triangle on ($x_t$, $y_t$) for the next three measurements which will give ($x_{t+\Delta t}$, $y_{t+\Delta t}$, $S_{t+\Delta t}$), and thus the track of the moving particle.

*Theoretical capabilities*

The maximum speed for the tracking of an individual NP is determined by the time needed to perform the three measurements. We use integration times of 5 ms per point separated by 6 ms waiting times to ensure stability of the piezoscanner and data transfer. Consequently, the tracking rate of the position of nanoparticles is ~30 Hz. The maximum speed of a moving object that can be tracked is thus limited to that of NP which can not escape from the triangle during the 3 measurements. Using a triangulation radius of 180 nm (~1.5 × the width of a single NP LISNA profile(25)), objects moving with diffusion speeds up to ~0.2 μm²/s will be tracked.

*Tests on simulated trajectories.*

We first tested SNaPT by generating 2D Brownian movements of individual NPs embedded in thin polyvinyl alcohol (PVA) films with the piezoscanner stage (see Fig. 1). The tracking scheme was then used to recover the simulated (and known) trajectories of the NPs as a function of the input diffusion constant. The tests were conducted for different detection SNRs (30 to 100). From the recovered trajectories, three main parameters were calculated: the ratio of successful tracks, defined as tracks which were not lost before a given time limit (200 data points, 7.5 s), the instantaneous diffusion constant $D$ (see methods) and the deviation from the simulated trajectory. For SNRs ~ 30, diffraction limited beams and a triangulation radius of 180 nm, a cut-off in the diffusion constants is found around 0.15 μm²/s, above which only few simulated trajectories could be retrieved (Fig 2). Below this value, a very good agreement between the generated and measured diffusions is obtained over more than 2 orders of magnitude of D and the standard deviation between the measured and the simulated trajectories gives the pointing accuracy of the method: 20 nm for SNR~30 or 7 nm for SNR~100. Only few trajectories with high diffusion constants (0.15-0.3 μm²/s) could also be retrieved. Interestingly, by enlarging the heating beam size and the triangulation radius by 50%, the cut-off in the diffusion constants could be increased by ~30% (not shown) but at the price of a 10% loss of the pointing accuracy. Measurements of very slow diffusion constants are limited by the pointing accuracy and the stability of the setup. The minimum detectable diffusion constant $D_{min}$ was experimentally determined with SNR~30 to be equal to $D_{min}=1.4\times10^{-3}$ µm²/s (Fig. 2).

### *Tracking on live cells*

For applications on live cells, we validate the SNaPT technique by measurements of the well characterized lateral diffusion of mGlurR5a (a metabotropic glutamate receptor, member of the super-family of G-Protein coupled receptors) in the plasma membrane of live COS7 cells (Fig 3 (a))(28). The cells were transfected and expressed a mGlurR5a-myc cDNA fusion construct consisting of the mGluR5a receptor sequence and a N-terminal extracellular myc tag. Those proteins were first sparsely labeled with fluorescent antibodies (anti-myc-Cy5) in order to perform SMT control experiments. For SNaPT, a secondary labeling stage, with either antiIgG-10nm or antiIgG-5nm gold antibodies was used (see methods). LISNA images of non-labeled COS7 cells show intrinsic signal only at regions close to the nucleus(29). Apart from these regions which will not be considered in this work, a 2 fold increase of the noise is found and no signal is detected. Transfected cells are easily discriminated from untransfected ones in fluorescence images (not shown). Specificity of the gold labeling was ensured as LISNA images of stained non-transfected cells are similar to that of non-labeled cells.

LISNA images of cells with labeled receptors reveal the presence of point-like signals which correspond to immobile (or slowly moving) mGluR5-linked NPs. But most NPs move during the raster scan of the sample, which produces characteristic stripe signals (see fig 3(c)). Fig 3(c-d) further shows an example of 2 SNaPT trajectories (20s acquisition time each, see supplementary materials for trajectories of several minutes). Rapid changes in diffusion regimes are clearly visible as reported previously on PtK2 cells and neurons(28). Although the signal exhibits increased fluctuations when the NP is in a fast diffusive state, it remains relatively stable during the overall trajectory (Fig. 3e) allowing a precise assignment of the number of tracked NPs. Fig 3(b) presents the histogram of the signals measured during 48 SNaPT recordings of mGlurR5a receptors labeled with 5nm gold NPs. The distribution contains two main peaks corresponding to the detection of one NP in the first peak and two NPs in the second(22).

Fig. 4(a-c) shows the distributions of *D* measured on the first 750ms of trajectories acquired by SMT and by SNaPT (with 5nm and 10nm gold NPs labeling). The median of the three distributions were not significantly different, validating the SNaPT method on the time scales accessible by SMT (p<0.05 Mann-Whitney-Wilcoxon test).

We will now address the important issue of the possible effect of the laser intensities used in SNaPT (200-400 kW/cm²). First, the local temperature rise due to laser absorption in the vicinity of the NP can be estimated. Since the thermal conductivity of metals is much higher than that of the surrounding medium, the temperature inside a spherical NP can be considered uniform and equal to the temperature at its surface. It writes: $T_{surf} = \frac{\sigma_{abs} I}{4\pi \kappa a}$ where *I* is the heating intensity, $\sigma_{abs}$ the NP absorption cross section, $\kappa$ the thermal conductivity of the medium (water) and *a* the radius of the NP. For 5 nm gold NPs in aqueous medium, and $I = 400$ kW/cm² one finds a rather low NP temperature rise of ~ 1.5 K. Furthermore it decreases as the inverse of the distance from the NP surface. Second on short time scales, SMT and SNaPT give the same results despite different excitation intensities (<10 kW/cm² for SMT). Third, we investigated longer time scales by recording 115 trajectories (600 data points each, 15 cells) and calculating $D$ over a sliding window (25 data points) along these trajectories (Fig.4(d)). No evolution of the mean and standard deviation of the distributions of $D$ is noticed and the proportion of immobile NPs do not change. Finally, 2×2 μm² regions of cells were exposed during 30 minutes to 400 kW/cm² continuous laser illumination. Comparison of images taken before and after the illuminations revealed no discernible adverse effect on the cell morphology as compared to control samples (Fig.4(e) (f)). Altogether these data indicate that SNaPT should allow single molecule experiments on live cells when long recordings and nanometer-sized labels are needed.

*SNaPT on live neurons*

For future applications, we further tested SNaPT by recording long trajectories of diffusing AMPA receptors (AMPARs) in the membrane of live neurons. The first evidence for lateral motion of AMPARs came from SPT studies where AMPARs labeled with latex bead (~0.5 μm diameter) were tracked at the surface of cultured hippocampal neurons(6). Using SMT experiments and thus reduced label sizes, Tardin et al(13) showed that AMPARs can also be mobile inside synapses. Altogether these studies established that AMPARs alternate between different membrane compartments through lateral diffusion.

The measure of the dynamics of these exchanges is important for the understanding of the synaptic physiology(30). As SMT is limited to short acquisitions times, the need for methods like SNaPT to follow nanometer sized labels at video rate for long times is required.

We thus labeled a small proportion of surface expressed native AMPARs containing the GluR2 subunit in live cultured hippocampal neurons through short incubations with antiGluR2 antibodies. As a secondary labeling stage, we used F(ab)-5nm gold conjugates (see methods). Similarly to mGluR5a imaging in COS7 cells, LISNA images of AMPARs on live neurons reveal the presence of immobile (point-like signals) and mobile objects (stripes, see Fig 5(a)). Intrinsic signals originating from some portions of neurites were also observed(29). We performed several recordings of single 5 nm gold NP-linked AMPARs for more than 5 minutes at video rate, corresponding to > 10000 data points (see movie in supplementary materials). Noteworthy and in comparison with imaging methods, SNaPT dramatically reduces data storage and data processing requirements for long trajectories since it directly measures the NPs position (and signal) as a function of time.

Similarly to previous SPT and SMT studies(6, 13, 14), GluR2 containing AMPARs are found in very diverse diffusion states by SNaPT (Fig 5(b)). In particular, fast diffusion time periods alternate with reduced and confined diffusion states (see inset of Fig 5(b)). These latter were previously assigned to AMPARs diffusing either in receptor clusters or inside synapses. For direct measurements of receptor dynamics and residency times inside synapses, SNaPT studies will require a further size reduction of the ligands between the receptor and the gold nanoparticle.

**Conclusion**

We have developed a new tracking method which combines the advantages of small marker size afforded by SMT and the unlimited observation time afforded by SPT with stable signals. In this work, we have only considered tracking of nanoparticles in the two dimensions of a planar membrane. The method itself, however, can be easily extended to full 3D tracking by simply adding a fourth measurement point to the presented scheme(27). We have demonstrated that SNaPT allows long recordings of membrane protein lateral diffusion with signal stabilities not yet achievable with fluorescent labels even by using semi-conductor nanocrystals. Further use of optimized bioconjugation strategies to bind the 5nm gold nanoparticles to the target protein with minimal ligand sizes(31), will make SNaPT a powerful approach for the study of individual proteins in vitro or in live cells.

## Methods

*LISNA setup*

We built a microscopy setup to perform SNaPT and wide-field SMT as well as white light imaging of cells (Figure 1). A non-resonant probe beam (HeNe, 632.8 nm) and an absorbed heating beam (532 nm, frequency doubled Nd:YAG laser) are overlaid and focused on the sample by means of a high NA microscope objective (100×, NA=1.4). The intensity of the heating beam is modulated at a frequency $\Omega$ (typically $\Omega/2\pi = 700$ kHz) by an acousto-optic modulator. A second microscope objective (80×, NA=0.8) collects the interfering probe-transmitted and forward-scattered fields. The intensity of the heating beam sent on the samples was 400 kW/cm$^2$ for 5 nm gold NPs and 200 kW/cm$^2$ for 10 nm gold NP. The forward interfering fields are collected on a fast photodiode and fed into a lock-in amplifier in order to extract the beat signal at $\Omega$. Integration times of 5 ms are used. Images and tracking are performed by moving the sample over the fixed laser spots by means of a 2D piezo-scanner.

*Fluorescence and white light imaging.*

Fluorescence images are recorded using the probe beam in a wide-field excitation configuration by inserting an additional lens to focus the beam in the back aperture of the objective lens. Samples were illuminated for 10 ms at a rate of 30 Hz on a surface of 400 μm$^2$ with an intensity of 7±1 kW/cm$^2$. The fluorescence from single Cy5 fluorophores is collected in epi and imaged on a back-illuminated thinned CCD camera. White light images are recorded by the same CCD camera using a standard condenser for illumination.

*Cell culture, transfection of COS7 cells and staining.*

COS7 cells were cultured in DMEM medium supplemented with streptomycin (100 µg/ml), penicillin (100 U/ml), and 10% bovine serum in a humidified atmosphere (95%) at 5% $CO_2$ and 37°C. Cells were used for 12-14 passages and were transferred every 4 days. For transfection the cells were plated onto 15 mm #1 glass plates to a confluence of ~ 30% and incubated with 1 µl FUGENE and 0.5 µg DNA coding for a metabotropic receptor for glutamate containing of myc-tag at the extracellular N-terminus (mGluR5a-myc(28)). Transfection efficiency was on the order of 40%. After 12h, immunostaining was performed using antimyc antibodies tagged with Cy5 dyes (herein named amyc-Cy5, 3 min at room temperature, 20 µg/ml, 0.3% BSA). After two rinses in PBS, a secondary immunostaining by antiIgG-10 nm or antiIgG-5 nm gold (goat anti-mouse, BBInternational, to label the amyc-Cy5, Auroprobes Amersham, 3 min at room temperature, 0.3% BSA) was performed at a antibody concentration of 300 ng/ml, followed by 3 rinses in medium. The coverslips were then mounted in a custom chamber with culture medium supplemented with 20 mM Hepes. All data were taken at room temperature within 20 minutes after a last rinse.

*Neuron culture, GluR2 staining*

Hippocampal neurons from 18 days old rat embryos were cultured on glass coverslips as previously(6). For SNaPT experiments, 7-10 DIV neurons were incubated 3 min at room temperature with 10 µg/ml anti-GluR2. After two rinses in culture medium, a secondary immunostaining by F(ab)-5nm gold conjugates (5nm gold conjugate goat F(ab')2 anti-mouse

IgG, BBInternational, to label the anti-GluR2, Auroprobes Amersham, 3 min at room temperature, 0.3% BSA) was performed at a concentration of 300 ng/ml followed by 3 rinses in culture medium. After fast rinses, the coverslips were mounted in a custom chamber with culture medium supplemented with 20 mM Hepes. All data were taken at room temperature within 20 minutes after the last rinse.

*Data analysis*

In the case of SMT, trajectories are recorded following the procedure described in(13). Only SMT trajectories containing at least 25 data points were analyzed. For each track (recorded by SNaPT or SMT), the Mean Square Displacements (MSDs) are calculated on 25 consecutive data points (~1 s) along the trajectory and the instantaneous diffusion constant $D$ is deduced as a function of time, from the initial slope of each MSD curve (fitted on the first 7 points).

## Acknowledgements

We thank Pierre Gonzales for COS7 cell cultures, Christelle Breillat and Delphine Bouchet-Teissier for neuron cultures. G.A. Blab acknowledges financial support by "Fonds zur Förderung der wissenschaftlichen Forschung" (FWF, Austria, Schrödinger-Stipendium) and the "Fondation pour la Recherche Médicale" (FRM, France). This research was funded by CNRS (ACI Nanoscience and DRAB), Région Aquitaine and the French Ministry for Education and Research (MENRT).

## Figures

Figure 1: Schematic of the experimental setup. Inset: LISNA image of 5 nm NPs embedded in a PVA film (SNR~30).

Figure 2: Tracking capabilities: ratio of successful tracking (open circles) and measured diffusion constant (squares with standard deviations) for generated 2D Brownian movements of individual NPs embedded in thin PVA film (SNR ~30). The arrow indicates the mean apparent diffusion constant found for immobile NPs, namely $4\times10^{-4}$ $\mu m^2/s$.

Figure 3: (a) Biological construct. (b) Histogram of the signals for 5 nm gold NPs detected on COS7 cells. (c) LISNA image of a portion of a transfected COS7 cell labeled with 5nm gold NPs detected with a SNR ~30. Red lines are trajectories recorded with single receptors diffusing on the cell membrane. (d) Zoom onto one recorded trajectory. (e) Time trace of the signal while tracking one of those receptors. Increased signal fluctuations correspond to the NP in a fast diffusive state.

Figure 4: (a) to (c) Distributions of the instantaneous diffusion constants $D$ of mGluR5a on COS7 cells, obtained with respectively SMT, SNaPT of 10 nm NPs and SNaPT of 5 nm NPs. The first bar (labelled with *) contains all the data for $D<D_{min}$. (d) Evolution of the distribution of instantaneous diffusion constants during long acquisitions (mean and standard deviation of the mobile fraction) (e) (f) White light images of a COS7 cell before (e) and after (f) 3 periods of 10 min continuous illumination over $2\times2\mu m^2$ areas (white frames on (e)).

Figure 5: White light image (a, left) and LISNA image (a, right) of a live neuron labelled with gold NPs. LISNA image exhibits signals from 2 moving and 1 stationary GluR2 receptors labelled with gold NPs (arrows). (b) Trajectory of an individual GluR2 receptor labelled with a 5nm gold NP (>5 minutes, 9158 data points) acquired at video rate on a live neuron (see

also a movie in supplementary materials).

Figure 1

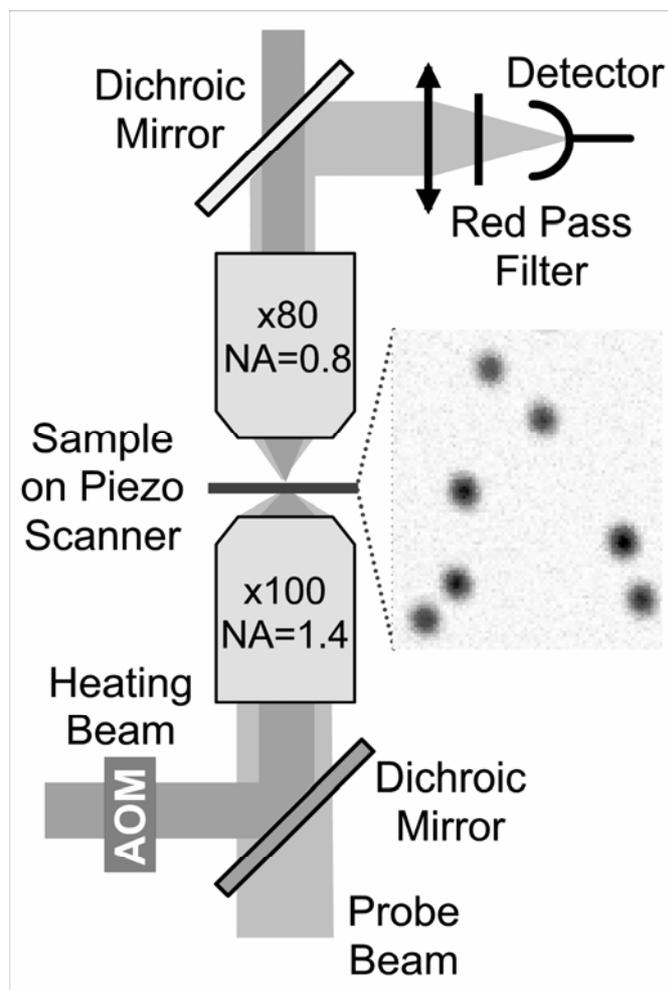

Figure 2

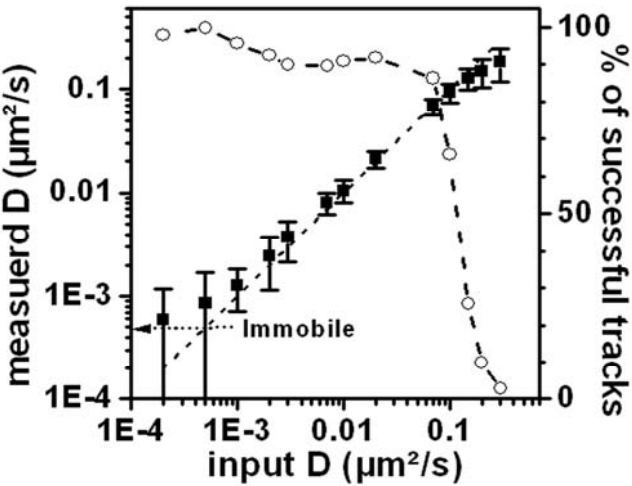

Figure 3

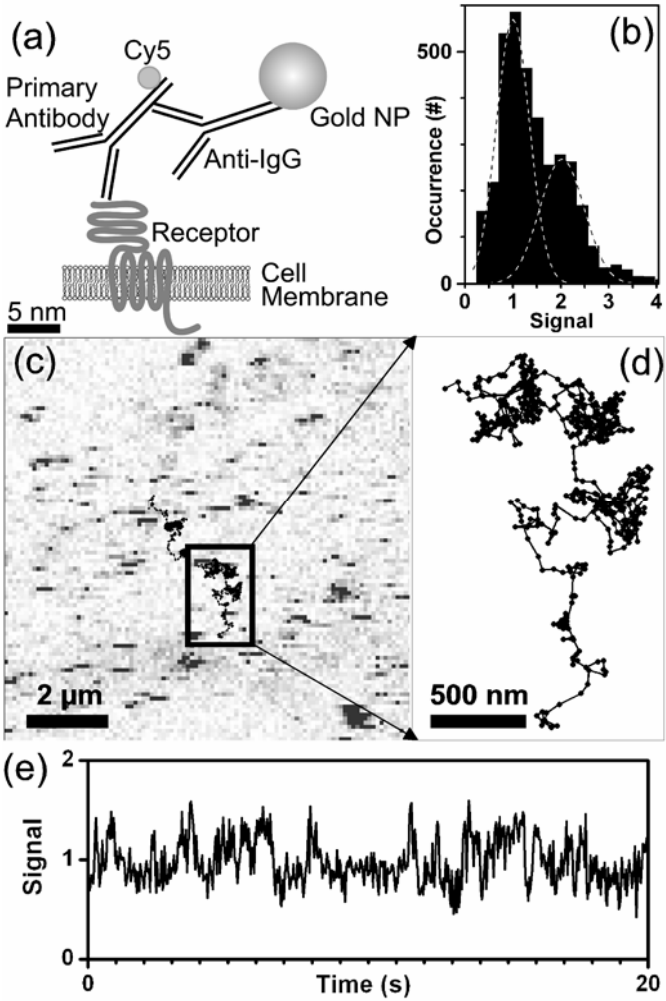

Figure 4

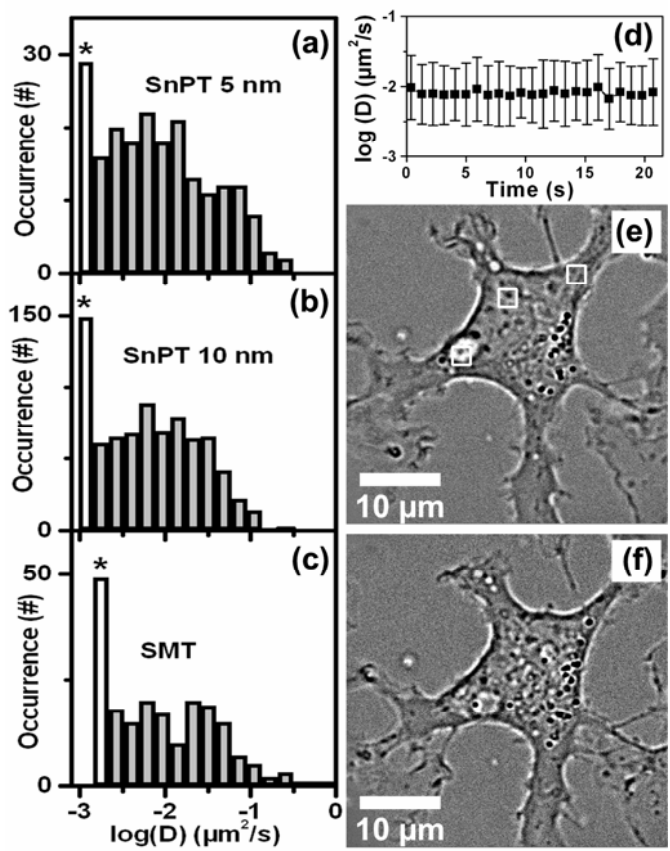

Figure 5

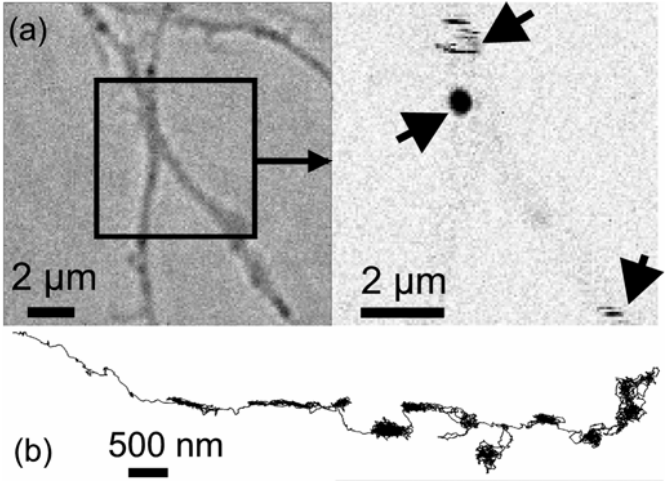